\documentclass[11pt]{article}

\usepackage{graphicx}
\usepackage{float}

\usepackage{amsmath}
\usepackage{a4wide}
\usepackage{bbm}
\usepackage{natbib}

\bibpunct[,]{(}{)}{,}{a}{}{,}

\begin{document}


\title{Fracture of complex metallic alloys:\\
An atomistic study of model systems}

\author{
{Frohmut R\"osch and Hans-Rainer Trebin}\\
{\normalsize Institut f\"ur Theoretische und Angewandte Physik,}
{\normalsize Universit\"at Stuttgart,}\\
{\normalsize Pfaffenwaldring 57, 70550 Stuttgart, Germany}\\ \\
{Peter Gumbsch}\\
{\normalsize Institut f\"ur Zuverl\"assigkeit von Bauteilen und Systemen,}\\
{\normalsize Universit\"at Karlsruhe, Kaiserstr.~12, 76131 Karlsruhe,
  Germany}\\
{\normalsize Fraunhofer Institut f\"ur Werkstoffmechanik,}
{\normalsize W\"ohlerstr.~11, 79108 Freiburg, Germany}
}

\date{}

\maketitle


\begin{abstract}

Molecular dynamics simulations of crack propagation are performed for two
extreme cases of complex metallic alloys (CMAs): In a model quasicrystal the
structure is determined by clusters of atoms, whereas the model C15 Laves
phase is a simple periodic stacking of a unit cell.
The simulations reveal that the basic building units of the structures also
govern their fracture behaviour. Atoms in the Laves phase play a comparable role
to the clusters in the quasicrystal. Although the latter are not rigid units,
they have to be regarded as significant physical entities.

\end{abstract}


\section{Introduction}\label{Intro}

Complex metallic alloys are intermetallic compounds with large unit
cells containing from tens up to thousands of atoms. Often, distinct local
arrangements of atoms -- clusters -- can be viewed as building units. Both, the
cluster diameter and the lattice constant imply length scales which
should be reflected in physical properties.
CMAs frequently combine interesting properties like high melting point, high
temperature strength, and low density. However, possible applications are
often limited by extreme brittleness at low or ambient temperature.
To enlighten the role of clusters and periodicity in fracture, we perform
molecular dynamics simulations of two extreme cases of CMAs: An icosahedral
quasicrystal and a C15 Laves phase.
As we are interested in the general qualitative features of the structures, we use
three-dimensional model systems consisting of about five million atoms and
model potentials (Lennard-Jones). This deliberate choice in the past often
helped to reveal fundamental aspects of fracture (see e.g.~\cite{Abraham2003}).
The quasicrystal can be viewed as a CMA with an infinitely large unit cell,
such that no periodicity is present and clusters are the main feature of the
structure. On the other hand, the C15 Laves phase has 24 atoms in the
cubic unit cell and no clusters. The structure of the Laves phases
is determined by periodicity but already quite complex, such that complicated
deformation mechanisms might emerge (see e.g.~\cite{Chisholm2005}).


\section{Models and method}

The three-dimensional model quasicrystal used in our numerical experiments has
been proposed by \cite{Henley1986} as a structure model for icosahedral
(Al,Zn)$_{63}$Mg$_{37}$. It is built up from the prolate and oblate
rhombohedra of the three-dimensional Penrose tiling. As we do not distinguish
between Al and Zn type atoms, we term its decoration icosahedral binary model
(for details see e.g.\ \cite{Roesch2005}). In the upcoming figures the Al and
Zn type atoms (A atoms) are displayed as grey (online: red) balls, whereas Mg
type atoms (B atoms) are shown in black (online: blue). The shortest distance
between two A atoms is denoted $r_{0}$ which corresponds to about 2.5 \AA.
Inherent in the structure are
Bergman-type clusters, which also can be viewed as basic building units.

As no reliable ``realistic'' effective potentials are available for the
fracture of CMAs, the interactions are modelled by simple
Lennard-Jones pair potentials (see Sec.~\ref{Intro}, \cite{Roesch2004}, and
\cite{Roesch2005}). These potentials keep the model stable even under strong
deformations or introduction of point defects and have been used in our group
together with the icosahedral binary model to simulate dislocation motion
(\cite{Schaaf2003}) and even shock waves (\cite{Roth2005}). Very similar
potentials have shown to stabilize the icosahedral atomic structure
(\cite{Roth1995}). By the choice of these model potentials and model
structures we qualitatively probe the influence of structural aspects of the
investigated compounds without being specific to a special kind of
material. Model potentials are often used in fracture simulations and have led
to useful insight into fundamental mechanisms (see
e.g.~\cite{Abraham2003}). The minimum of the Lennard-Jones potential for the
interactions between atoms of different kind is set to twice the value of that
for atoms of the same type. However, the conclusions drawn from our
simulations remain essentially unaffected by setting all binding energies
equal, which again indicates that we are mainly probing structural effects.

A fundamental building unit of the simulated quasicrystal -- the prolate
rhombohedron -- in a slightly deformed way forms the cubic C15 Laves phase
A$_{2}$B by periodic arrangement. But in this structure no clusters of the
quasicrystal are present. Because of the close structural relationship of the
C15 Laves phase to the quasicrystal model, we use the same model potentials.

Our samples have dimensions of approximately $450 r_{0} \times 150 r_{0}
\times 70 r_{0}$ and contain about five million atoms. Periodic boundary
conditions are applied in the direction parallel to the crack front.
To simulate mode I fracture, we first determine potential cleavage planes.
According to \cite{Griffith1921} crack propagation becomes possible, when the
elastic energy is sufficient to generate two new fracture
surfaces. Thus, potential cleavage planes should be those with low surface
energies. In the quasicrystal these are specific
twofold and fivefold planes (see \cite{Roesch2005}). There atomically sharp
seed cracks are inserted. The samples are uniaxially strained perpendicular to
the crack plane up to the Griffith load and then relaxed to obtain the
displacement field of the stable crack at zero temperature. Subsequently, a
temperature of about $10^{-4}$ of the melting temperature is applied. The
sample is further loaded by linear scaling of the corresponding
displacement field for this temperature. The response of the system then is
monitored by molecular dynamics techniques. The radiation emitted by the
propagating crack is damped away outside an elliptical region to prevent
reflections (see \cite{Roesch2005} and \cite{Gumbsch1997}).


\section{Results and discussion}

\begin{figure} [htb]
\centering
\includegraphics[width=1.0\linewidth,clip=]{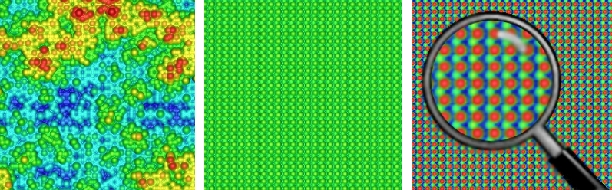}
\caption[ ]{Sections of typical fracture surfaces of an
  icosahedral model quasicrystal (left) and of a C15 model Laves phase
  (middle, right). The colour coding in the left and middle picture is
  adjusted to the cluster diameter (online: from blue to red). For details see
  text. The side length of the squares is about 14 nm.
\label{fig1}}
\end{figure}

\begin{figure} [htb]
\centering
\includegraphics[width=1.0\linewidth,clip=]{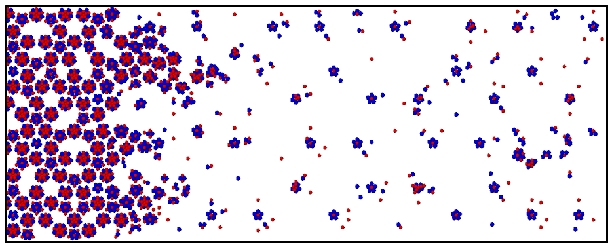}
\caption[ ]{Clusters cut by the crack in a model quasicrystal:
  Only the smaller parts of those clusters are displayed that were divided by
  the crack. Obviously, the dynamic crack (right) intersects fewer clusters
  than the low energy seed crack (left). The cleavage plane is located
  perpendicular to a fivefold axis, the crack propagated in twofold direction.
\label{fig2}}
\end{figure}

In the model quasicrystal we investigated how clusters and the plane
structure influence crack propagation. A paper on this detailed study has
recently been published (see \cite{Roesch2005}), where brittle fracture
without any crack tip plasticity was reported.
The following results from that paper indicate that the clusters determine
the brittle cleavage fracture of the model quasicrystal:
First, circumvention or intersection of clusters slows down the
cracks. Second, the fracture surfaces show characteristic height variations
giving rise to an overall roughness on the cluster scale (see Fig.\
\ref{fig1}, left). Lines along which the clusters are located are also visible
in the height profiles. Third, the crack intersects fewer clusters than a
planar cut with low surface energy does (see Fig.\ \ref{fig2}).
Another observation of the simulations is that the plane structure also
influences fracture. Cracks located perpendicular to twofold and fivefold axes
fluctuate about a constant height.
Thus, the roughness of the crack surfaces can be assigned to
the clusters, whereas constant average heights of the fracture surfaces
reflect the plane structure of the quasicrystal.

\begin{figure} [htb]
\centering
\includegraphics[width=0.85\linewidth,clip=]{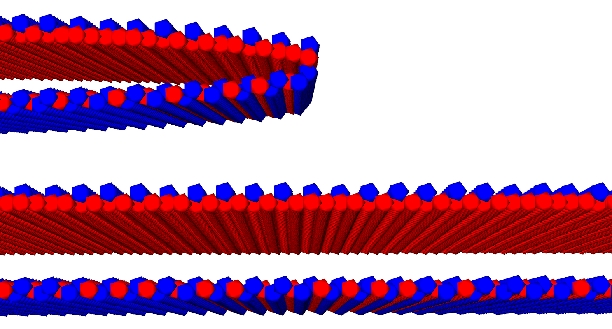}
\caption[ ]{View inside a crack of the C15 model Laves
  phase. Perfect brittle cleavage fracture is observed.
\label{fig3}}
\end{figure}

Now we compare simulation results of the quasicrystal to those of the C15
Laves phase. For this structure we also observe brittle failure. But the
fracture surfaces at low loads -- if at all -- only are rough on an atomic
scale. This becomes apparent from Figs.\ \ref{fig1} and \ref{fig3}. 

In Fig.\ \ref{fig3} only atoms near a (111) fracture surface are displayed.
The seed crack shown on the top propagated in $[2 \bar 1 \bar 1]$
direction. The material perfectly cleaved (Fig.\ \ref{fig3}, bottom), which
resulted in smooth fracture surfaces.

In Fig.\ \ref{fig1} sections of geometrically scanned fracture surfaces of the
icosahedral model quasicrystal (left) and of the C15 model Laves phase
(middle, right) are compared. The cleavage planes are located perpendicular to
a twofold (left) and a $[$010$]$ axis (middle, right). The crack propagated
along a twofold (left) and a $[$101$]$ axis (middle, right). The colour coding
in the left image is adjusted to the cluster diameter (online: from blue to
red). When colour coded like the quasicrystal, the fracture surfaces of the
C15 Laves phase lack any roughness (middle, online: only green). After
adjusting the colour coding, atomic rows become visible (right). So, fracture
surfaces of the quasicrystal are rough on the cluster scale, whereas those of
the Laves phase only are rough on an atomic scale.

\begin{figure} [htb]
\begin{center}
\begin{tabular}{cc}
\includegraphics[height=0.4\linewidth,angle=-90,clip=]{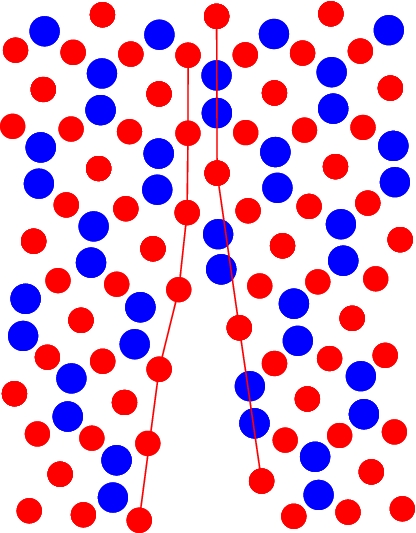}&
\includegraphics[height=0.4\linewidth,angle=-90,clip=]{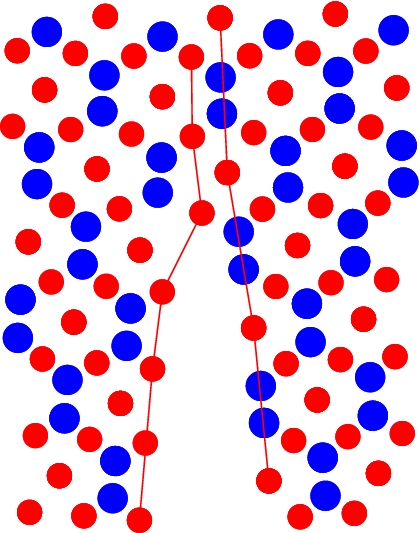}\\
a)&b)\\
\includegraphics[height=0.4\linewidth,angle=-90,clip=]{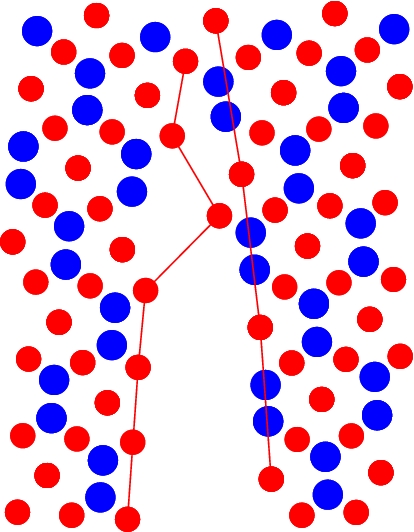}&
\includegraphics[height=0.4\linewidth,angle=-90,clip=]{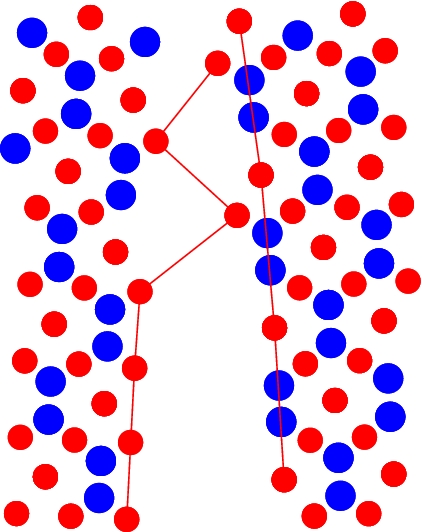}\\
c)&d)\\
\end{tabular}
\caption[ ]{Fracture of a C15 model Laves phase:
 Atomic configurations in the vicinity of a propagating crack (time sequence).
 The fracture surface is located perpendicular to the $[$011$]$ direction, the
 crack propagates along the $[$100$]$ direction (from left to right).
\label{fig4}}
\end{center}
\end{figure}

Thus, atoms in the Laves phase play a comparable role to the clusters in the
quasicrystal -- they determine the overall roughness of the fracture
surfaces. The atomistic view of cleavage on a (011) plane in Fig.\
\ref{fig4} reveals an interesting effect\footnote{A similar behaviour is
observed for the orientation shown in Fig.\ \ref{fig1} (middle, right).}:
If the seed crack there would be continued, the lines would terminate
the upper and lower halves of the sample. However, as can be seen in the time
sequence of Fig.\ \ref{fig4}, this
is not the case: The dynamic crack instead takes a zig-zag like route. Entire
atomic rows alternately move upwards and downwards. This leads to rather
symmetric upper and lower fracture surfaces, the creation of which also
requires a comparable amount of energy. This rather symmetrical creation of
fracture surfaces is favoured, even though a planar cut would lead to surfaces
with lower total energy. Thus, the actual fracture path cannot be predicted by
a simple energy criterion. Such a behaviour also was observed in B2 NiAl (see
e.g.\ \cite{Gumbsch2001}) and is a consequence of lattice trapping
(\cite{Thomson1971}), which -- similar to the Peierls barrier for
dislocation motion -- allows overloads for cracks that do not result in
crack propagation. The increased load for a crack to propagate will therefore
not necessarily lead to fracture surfaces of lowest energy. The discrete
nature of matter is responsible for these observations. The fracture path
is strongly influenced by the arrangement of atoms near the crack front, which
depends on the initial cleavage plane as well as the crack propagation {\em
  direction}.


\section{Conclusions and outlook}

In conclusion, the simulation results of the two extreme cases of model CMAs
indicate that the basic building units of the structures govern also their
physical properties. The role of atoms in the Laves phase is -- from a certain
point of view -- played by the clusters in the quasicrystal. Although these
are not rigid units, they are significant physical entities.

To further enlighten the fracture processes in the C15 Laves phase, we
currently perform simulations with force-matched (\cite{Ercolessi1994},
\cite{BrommerICQ9}) effective embedded atom method potentials for NbCr$_{2}$
(\cite{Roeschtbp}). Although we expect that the overall qualitative behaviour
(e.g.\ the roughness of the fracture surfaces) is already represented well by
our simple model potentials, results certainly will differ quantitatively for
diverse materials, i.e.\ interactions. Especially, the lattice trapping
mentioned above strongly depends on the potentials used.

Future studies will concentrate on material specific simulations on systems of
CMAs, in which compounds with very different unit cell sizes and local
arrangements exist. New effects are expected e.g.\ when the size of the unit
cell and cluster diameters become comparable.


\section*{Acknowledgement}

Financial support from the Deutsche Forschungsgemeinschaft under contract
number TR 154/20-1 is gratefully acknowledged.


\addcontentsline{toc}{chapter}{References}
\bibliographystyle{abbrvnat}


\vspace{1cm}

\section*{Figures}

\end{document}